\documentclass{article}
\usepackage{ijcai26}

\usepackage{graphicx}
\usepackage{subcaption} 

\usepackage{times}  
\usepackage{helvet}  
\usepackage{courier}  
\usepackage[hyphens]{url}  
\usepackage{graphicx} 
\usepackage{natbib}  
\usepackage{caption} 
\usepackage{amsmath}
\usepackage{amsthm}           
\usepackage{lineno} 
\theoremstyle{definition}     
\newtheorem{definition}{Definition}
\usepackage{amssymb}
\usepackage{amsfonts}
\usepackage{booktabs}   
\usepackage{comment}
\usepackage{placeins}

\usepackage{algorithm}
\usepackage{algorithmic}

\usepackage{newfloat}
\usepackage{listings}
\DeclareCaptionStyle{ruled}{labelfont=normalfont,labelsep=colon,strut=off} 
\floatstyle{ruled}
\newfloat{listing}{tb}{lst}{}
\floatname{listing}{Listing}
\title{Why Study Emergent Behavior When You Can Regulate It? Aligning Multi-Agent Systems with Reward Prediction}

\author{
    Assaf Caftory\textsuperscript{1}\thanks{assafcaf@gmail.com},
    Almog Zemach\textsuperscript{1}\thanks{almog.zemach@post.runi.ac.il},
    Moshe Butman\textsuperscript{2}\thanks{moshe.butman@gmail.com},
    Doron Friedman\textsuperscript{3}\thanks{Corresponding author: doronf@runi.ac.il}\\
    \textsuperscript{1}Efi Arazi School of Computer Science, Reichman University, Herzliya, Israel\\
    \textsuperscript{2}Faculty of Computer Science, The College of Management Academic Studies, Rishon LeZion, Israel\\
    \textsuperscript{3}School of Communication, Reichman University, Herzliya, Israel
}

\usepackage{bibentry}

\begin{document}

\maketitle

\begin{abstract}

Multi-agent simulations are widely used to study complex social and ecological systems, where rich and often unexpected emergent behaviors arise from local interactions. A large body of prior work has focused on analyzing such emergent dynamics across domains. In this paper, we move beyond analyzing emergent behavior and introduce a learning-based mechanism for actively shaping it via social reward modeling. We introduce Multi-Agent Reward Prediction (MARP), a simple framework that extends preference-based reward modeling to multi-agent reinforcement learning. While the framework is designed to be applicable across multi-agent settings, the present empirical validation is limited to a single environment, and we therefore present MARP as a proof of concept within the studied domain. Rather than relying on handcrafted rewards, MARP learns a shared reward model from episode-level evaluations of collective outcomes, enabling decentralized agents to align their behavior with global social objectives.

We study MARP in the Harvest Game, a canonical sequential social dilemma modeling common-pool resource management and related real-world challenges. Our results show that MARP can be tuned to produce behavior that is more closely aligned with target social metrics than standard reward-based baselines, while the learned reward model captures subtle environmental structure without explicit programming. Crucially, MARP supports multiple and composite social objectives within a single training regime. By modifying only the high-level evaluation metric, the same framework seamlessly aligns agent behavior with diverse goals, including sustainability, equality, and peace, as well as combinations of individual and group-level objectives. These findings demonstrate that emergent multi-agent behavior can be treated not only as a phenomenon to study, but as a target of principled, data-driven regulation.

\end{abstract}

\section{Introduction}

Multi-agent simulations and agent-based models are widely used to study complex social, economic, and ecological systems, as simple local interaction rules can give rise to rich and often surprising emergent behavior. Increasingly, agents in such models learn and adapt via multi-agent reinforcement learning (MARL), which substantially expands the expressive power of agent-based models and supports applications across diverse domains, including economic forecasting \citep{axtell2025agent}, pandemic modeling \citep{kerr2021covasim}, and cooperation in social dilemmas \citep{axelrod1997advancing}.

Traditionally, such simulations have played a primarily explanatory role: researchers specify bottom-up rules, observe the resulting collective dynamics, and analyze how large-scale patterns such as cooperation, conflict, or collapse emerge. In this work, we argue for a complementary shift in perspective. Rather than treating emergence as something to be merely observed, we propose to systematically regulate multi-agent systems by introducing learning-based, top-down constraints on collective behavior. By using reinforcement learning (RL) to shape agent incentives according to global objectives, emergent dynamics can be steered toward desired outcomes rather than left to unfold uncontrolled.

This capability is directly relevant for engineering artificial multi-agent systems, where safety, sustainability, and coordination are central concerns. More broadly, methods for regulating populations of artificial agents also provide a formal lens for studying how collective behavior can be influenced in human societies, raising both practical opportunities and ethical questions about control, alignment, and governance.

Manually regulating collective behavior in multi-agent systems via reward design is notoriously brittle. Social objectives are typically defined at the group level rather than the individual agent, and depend on long-term, aggregate outcomes such as sustainability, fairness, or stability. These objectives are often delayed, non-Markovian, and only observable at the end of an episode, while agents act under local, partial observations and short-term feedback. Translating such global goals into dense, step-wise reward signals therefore requires extensive hand-engineering and strong assumptions about how individual actions relate to collective outcomes. In practice, even small reward mis-specifications can lead to unintended equilibria, reward hacking, or coordination collapse, especially as agents learn to exploit the proxy rather than the intended objective.

Handcrafted reward shaping embeds strong, often implicit assumptions about how individual actions aggregate into desirable collective outcomes. These assumptions are difficult to justify a priori and are tightly coupled to a specific environment, population size, or behavioral regime. As a result, reward functions that succeed in one setting often fail to transfer or generalize when agents, dynamics, or objectives change. The problem is compounded when multiple social objectives are involved. Supporting different or competing goals, such as efficiency, fairness, safety, or peace, typically requires redesigning the reward function and re-balancing terms for each objective. This makes manual reward design labor-intensive and fragile, limiting its usefulness as a systematic tool for regulating emergent behavior in complex multi-agent systems.

An alternative to manual reward design has emerged in single-agent RL through reward modeling and preference-based learning. Rather than specifying objectives as step-wise reward functions, this paradigm learns a reward model from high-level evaluations of behavior, such as trajectory comparisons or episodic outcomes \citep{christiano2017deep, stiennon2020learning, leike2018scalable}. The central idea is that supervision is provided at the level where goals are naturally defined, while learning operates at the level where control is required. A learned reward model thus acts as an intermediate representation, converting sparse, holistic judgments about overall performance into dense, temporally localized learning signals. This decoupling between objective specification and action-level optimization has proven effective for aligning behavior with complex, implicit goals, and is increasingly viewed as a core mechanism for addressing alignment in single-agent systems \citep{ngo2022alignment, ouyang2022training}.

Despite this progress, existing work lacks a systematic treatment of normative alignment in multi-agent systems. Most alignment-oriented MARL approaches regulate behavior by injecting handcrafted social preferences, intrinsic motivations \citep{jaques2019social}, or auxiliary rewards, such as fairness penalties, inequity aversion, or influence bonuses \citep{hughes2018inequity}. While effective in specific settings, these methods still rely on manual design choices that encode strong assumptions about desired behavior and often entangle objectives with environment dynamics. More recent preference-based MARL approaches move away from explicit shaping, but primarily focus on reconstructing environment rewards or improving task performance in cooperative benchmarks. As a result, preferences are treated as a means to optimize a predefined task, rather than as a mechanism for aligning emergent group behavior with explicitly normative, system-level objectives. The problem of learning to regulate what kinds of collective behavior emerge, rather than how efficiently a task is solved, therefore remains largely underexplored.

This work asks a simple but fundamental question: can preference-based reward modeling be used to regulate emergent behavior in multi-agent systems toward explicit social objectives? Rather than treating preferences as a proxy for unknown environment rewards or task performance, we examine whether high-level evaluations of collective outcomes can be translated into learning signals that systematically steer decentralized agents toward desired group-level norms, such as sustainability, equality, or peace. Framed this way, the goal is not to eliminate emergence, but to shape it, preserving the generative richness of multi-agent dynamics while aligning long-run outcomes with normative objectives defined at the system level.

To address this question, we introduce Multi-Agent Reward Prediction (MARP), a framework for regulating emergent behavior in decentralized multi-agent systems through learned rewards. MARP centers on a shared reward model trained from episode-level evaluations of collective outcomes, rather than step-wise or agent-specific supervision. Social objectives are specified only at the level where they naturally arise, namely as holistic judgments of group behavior over complete interactions. During learning, decentralized agents act on local observations and independently optimize rewards predicted by this shared model, without access to the global objective itself. In this way, MARP treats emergent behavior not as a byproduct to be analyzed after the fact, but as the object of optimization. Regulation emerges through learning and adaptation rather than explicit enforcement, hard constraints, or punitive mechanisms, allowing coordinated norms to arise organically from local decision making while remaining aligned with global social objectives.

We evaluate MARP in the context of sequential social dilemmas (SSDs), a canonical class of multi-agent environments in which individually rational behavior conflicts with long-term group welfare. In SSDs, short-term incentives encourage over-exploitation of shared resources or conflict, while sustainable outcomes require the emergence of norms such as restraint, coordination, and mutual accommodation. These environments are well suited for studying normative alignment because desirable behavior cannot be reduced to task completion, but instead depends on population-level patterns unfolding over time. Crucially, SSDs admit clear and interpretable social metrics, such as sustainability, equality, efficiency, and peace, which capture different dimensions of collective success without prescribing specific behaviors. SSDs therefore serve not merely as benchmarks, but as conceptual stress tests for alignment methods, exposing the gap between local optimization and global norms.

A central technical challenge is the mismatch between how supervision is provided and how control is exercised. Success signals are global and episodic, expressed as evaluations of entire multi-agent interactions rather than individual actions, while decision making is local, partial, and step-wise. Each agent observes only a limited view of the environment and acts without direct access to the collective objective, making credit assignment implicit and indirect. The learning system must infer how locally observed actions contribute to long-term, population-level outcomes in the absence of explicit intermediate feedback. Bridging this gap between global evaluation and local control is the core technical challenge addressed by MARP.

\emph{Contribution}. We introduce a simple framework for regulating emergent behavior in MARL by learning a shared reward model from episode-level social evaluations. The method is designed to be domain-agnostic, while the present validation is confined to the Harvest Game and should therefore be read as a proof of concept rather than a fully general result. We show that global, episodic preference signals are sufficient to induce locally meaningful reward predictions that guide decentralized agents, without architectural complexity, explicit coordination channels, or hand-engineered cooperation mechanisms. We further introduce and systematically compare two reward inference strategies, Local-Trajectory Inference, which learns from individual agent trajectories under global preferences, and Joint-Episode Inference, which conditions reward prediction on aggregated multi-agent episode representations. Empirically, we demonstrate in a sequential social dilemma that both approaches can reliably steer emergent behavior toward multiple and composite social objectives, including efficiency, sustainability, equality, and peace, outperforming standard reward-based baselines. Through analysis of the learned reward functions, we show that the model captures both environmental structure and social context, giving rise to stable and interpretable cooperative norms such as voluntary restraint under scarcity.

\footnote{Code is available at \url{https://github.com/almogze/marp}.}

\section{Background}


MARL extends single-agent RL to settings in which multiple agents learn and act simultaneously, each adapting its policy in response to both the environment and other agents. A central design axis in MARL is the degree of coordination between agents during learning. Centralized or partially centralized training schemes leverage shared information to stabilize learning, while decentralized approaches rely on local observations and independent updates, often leading to richer but harder-to-control emergent dynamics \citep{gronauer2022multi}.  These challenges are particularly acute in social dilemma settings, where individually optimal policies may conflict with desirable collective outcomes.

Social dilemmas capture settings in which individually rational behavior conflicts with collective welfare. In such environments, agents face incentives to maximize short-term personal gain, even when this leads to inefficient or harmful group-level outcomes. SSDs extend this structure to temporally extended interactions, where agents must balance immediate rewards against long-term sustainability and social stability \citep{leibo2017multi}.

SSDs reflect a common failure mode of decentralized learning: agents that optimize locally can converge to undesirable emergent dynamics, such as over-exploitation of shared resources or persistent conflict. The long-term and interactive nature of SSDs makes outcomes highly sensitive to policy interactions, highlighting the need for mechanisms that can shape emergent behavior toward socially desirable equilibria.

Prior work has shown that explicitly incorporating social incentives can improve cooperation in SSDs. For example, inequity aversion augments individual rewards with fairness-based penalties, reducing payoff disparities and promoting coordinated behavior \citep{hughes2018inequity}. Other approaches introduce intrinsic rewards based on an agent’s causal influence on others’ future actions, encouraging cooperative dynamics through social impact awareness \citep{jaques2019social}. 

\begin{figure}[htbp]
  \centering
  \includegraphics[width=8cm]{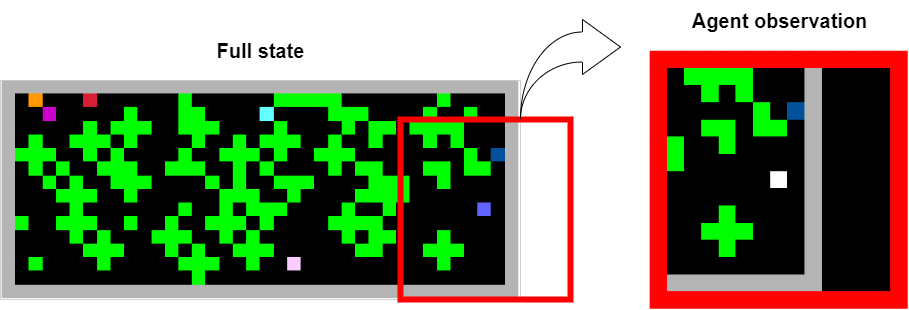}
  \caption{(Left) Full environment state showing green apples, gray walls, and agents as colored squares; the red box denotes the observation window of a single agent. (Right) The corresponding egocentric view, with the observing agent highlighted in white.}
  \label{fig:apples}  
\end{figure}

The commons game~\citep{SSDOpenSource, perolat2017multi} is a multi-agent environment modeling common-pool resource appropriation. Agents harvest renewable resources (``apples'') that regrow at density-dependent rates and may tag other agents to temporarily remove them from the map, without receiving direct reward for tagging (Figure~\ref{fig:apples}). This structure induces a social dilemma: aggressive harvesting and conflict can benefit individuals in the short term, while over-exploitation and excessive tagging reduce long-term collective productivity.

Typical learning dynamics progress through three phases observed in prior work: an initial exploration phase with inefficient harvesting, a tragedy phase characterized by resource depletion and conflict, and a maturity phase in which agents adopt more sustainable harvesting strategies that balance immediate reward with preservation of the commons.

Following \citet{perolat2017multi}, we evaluate behavior using several social metrics: \textit{Efficiency} (\(U\)), measuring total reward across agents; \textit{Equality} (\(E\)), computed via the Gini coefficient; \textit{Sustainability} (\(S\)), reflecting the temporal distribution of rewards; and \textit{Peace} (\(P\)), defined as the proportion of time steps in which agents are not tagged.

In the equations below, $N$ denotes the number of agents and $T$ the episode length; $R^i$ is the cumulative reward of agent $i$ over the episode, $r_t^i$ its reward at time $t$, and $o_t^i$ its observation at time $t$.

{\small
\begin{align}
U &= \mathbb{E}\left[\frac{1}{T} \sum_{i=1}^{N} R^i\right] \\
E &= 1 - \frac{\sum_{i=1}^{N} \sum_{j=1}^{N} |R^i - R^j|}{2N \sum_{i=1}^{N} R^i} \\
S &= \mathbb{E}\left[\frac{1}{N} \sum_{i=1}^{N} t^i\right], \quad t^i = \mathbb{E}[t \mid r_t^i > 0] \\
I(o) &=
\begin{cases}
1 & \text{if } o = \text{time-out observation} \\
0 & \text{otherwise}
\end{cases} \\
P &= \frac{\mathbb{E}\left[NT - \sum_{i=1}^{N} \sum_{t=1}^{T} I(o_t^i)\right]}{NT}
\end{align}
}

\section{Related Work}

\citet{zhu2024decoding} propose MAPT, a Transformer-based model that reconstructs dense multi-agent reward functions from trajectory preferences to approximate environmental rewards in complex benchmarks such as SMAC and Football. In contrast, MARP does not seek to infer environment rewards, but aligns emergent behavior with explicit episode-level social metrics, focusing on collective dynamics in social dilemmas rather than fine-grained control.

The most closely related line of work is the \textit{AI Economist} series, which introduced a two-level RL framework for policy design in agent-based simulations. In this paradigm, self-interested agents learn economic behavior while a centralized social planner simultaneously learns policy levers, most notably dynamic tax schedules, to optimize a specified social welfare objective, explicitly addressing the non-stationarity arising from planner–agent co-adaptation \citep{zheng2022aieconomist}. This work shows that learned policies can recover known theoretical solutions in simplified settings and improve equality–productivity trade-offs in richer spatiotemporal economies such as Gather--Trade--Build, giving rise to emergent phenomena including specialization and strategic responses to taxation (``tax gaming''). Subsequent extensions broaden the framework beyond taxation to data-driven and interpretable public-policy optimization in calibrated simulations, including pandemic policy and subsidy design, with an emphasis on robustness and explainability \citep{trott2021building}.

Our setting departs in several ways. First, rather than learning a planner policy that intervenes in the environment, we learn a shared reward predictor from preference comparisons over collective outcomes and use it to provide dense learning signals to decentralized agents. The learning problem we address is therefore one of multi-agent credit assignment under a non-Markovian global objective via reward modeling, rather than stabilizing two-level RL through curricula and regularization. As a result, while both lines of work aim to shape collective outcomes in multi-agent simulations, \textit{AI Economist} does so through explicit centralized mechanism design, whereas MARP operates through preference-based reward modeling that can be reused across objectives without introducing new environment-level interventions.

\section{Method}

\subsection{Settings and definitions}

\label{subsec:definitions}

We consider a multi-agent system where $n$ agents interact with an environment in discrete timesteps. At time $t$, agent $i$ receives observation $o_t^{(i)} \in \mathcal{O}^{(i)}$ and takes action $a_t^{(i)} \in \mathcal{A}^{(i)}$. Unlike conventional MARL where rewards come directly from the environment, in MARP agents receive rewards $\hat{r}_t^{(i)}$ from a learned reward model parameterized by $\theta$.
\begin{definition}[Trajectory]
A trajectory $\tau^{(i)}$ for agent $i$ is a sequence of observation-action pairs from $t_0$ to a terminal state $T$: $\tau^{(i)} = \langle (o_t^{(i)}, a_t^{(i)}) \rangle_{t=0}^{T}$.
\end{definition}

\begin{definition}[Episode]
An episode $\omega$ is the set of all agent trajectories in a multi-agent interaction: $\omega = \{\tau^{(1)}, \ldots, \tau^{(n)}\}$, where $n$ is the number of agents.
\end{definition}

\begin{definition}[Social Metric]
A social metric is a function $\phi : \omega \rightarrow \mathbb{R}$ that quantifies a group behavior from an episode.
\end{definition}

\begin{definition}[Preference Oracle]
A preference oracle $\mathbb{O}$ ranks episodes based on social metrics: $\omega_1 \succ \omega_2$ denotes that $\phi(\omega_1) > \phi(\omega_2)$. We define:

\textit{Episode-level oracle}: $\mathbb{O}_\omega: \omega_1 \times \omega_2 \rightarrow \{0,1\}$, where $\mathbb{O}_\omega(\omega_1, \omega_2) = 1$ iff $\omega_1 \succ \omega_2$.

\textit{Trajectory-level oracle}: $\mathbb{O}_\tau: \tau^{(i)}_{\omega_1} \times \tau^{(j)}_{\omega_2} \rightarrow \{0,1\}$, with $\tau^{(i)}_{\omega_1} \in \omega_1$, $\tau^{(j)}_{\omega_2} \in \omega_2$, and $\mathbb{O}_\tau(\tau^{(i)}_{\omega_1}, \tau^{(j)}_{\omega_2}) = 1$ iff $\omega_1 \succ \omega_2$.
\end{definition}

In our setup exact draws between trajectories are rare, and they are discarded. 

\subsection{Reward Model Design}
\label{subsec:rewardFucntion}

There are several decision points and alternatives for multi-agent reward prediction. Here we focus on a centralized local-view reward model, a single network shared among all agents. Evaluating additional architectures is left for future work. This approach introduces an interesting challenge: it requires reconciling individual agent inputs $\tau^{(i)}_{\omega}$ with group-level social metrics $\phi(\omega)$ used for training, creating an inherent information discrepancy. The model only receives local egocentric trajectories (partial observations) while the oracle judges entire multi-agent episodes (fully observable), so the learning signal is both global and sparse, and the model must infer how individual actions contribute to collective outcomes despite this inherent information asymmetry.

To address this challenge, we explore two training strategies. The first approach is \textbf{Joint-Episode} preference learning, where the reward model is trained on concatenated inputs that combine all agent trajectories from a given episode into a single sequence. Preferences over these aggregated inputs are provided by the episode-level oracle $\mathbb{O}_\omega$.

In contrast, \textbf{Local-Trajectory} preference learning trains the reward model on individual agent trajectories $\tau^{(i)}_{\omega}$, provided from each agent’s specific point of view, such as a very limited egocentric map window. Preferences are supplied by the trajectory-level oracle $\mathbb{O}{\tau}$. Using this method, the reward model architecture is based on single-agent input, with one observation-action pair as input and a single reward scalar as output. However, the reward model is simultaneously trained by all agents. 

\begin{figure}[htbp]
  \centering
  \includegraphics[width=\columnwidth]{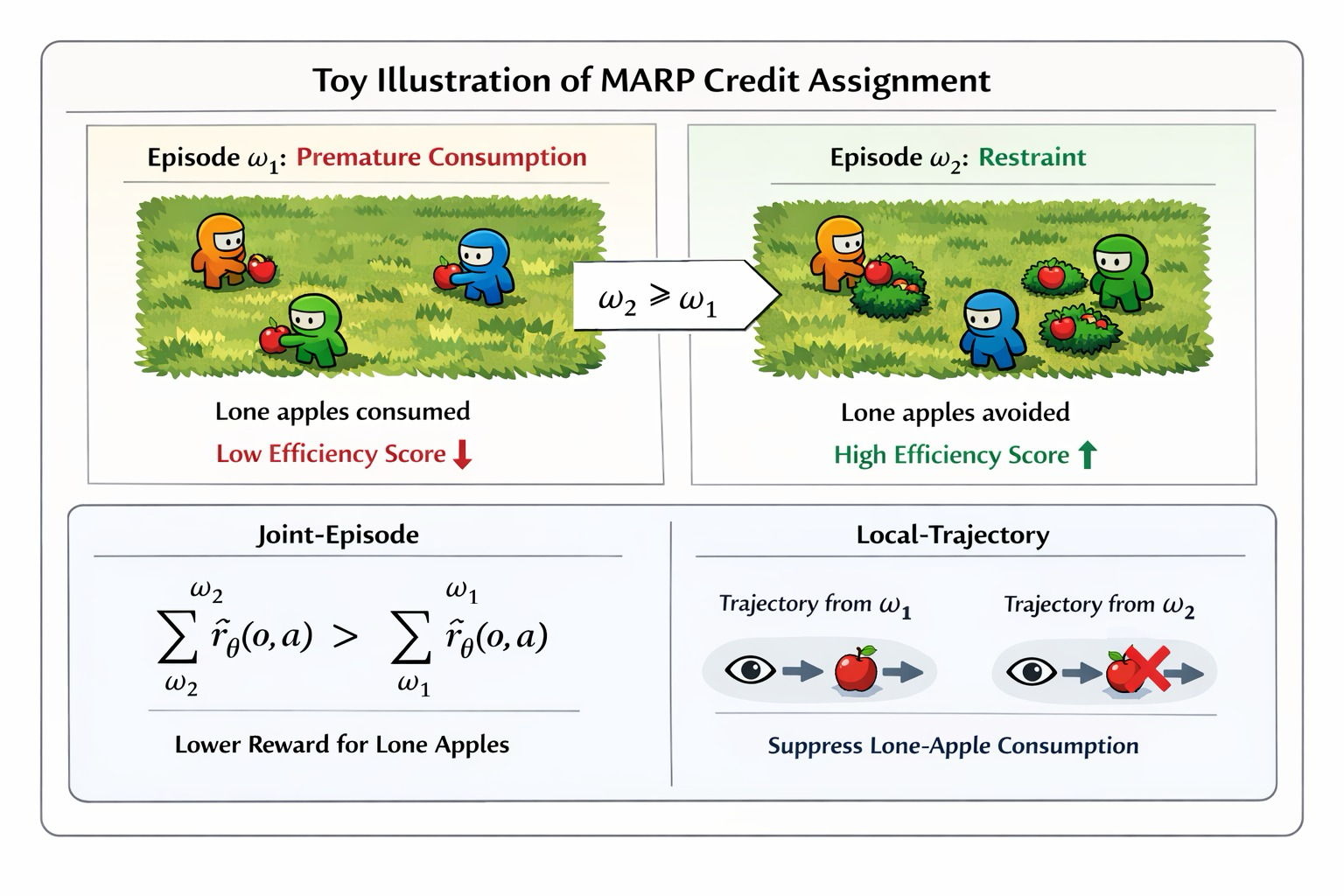}
  \caption{Toy illustration of MARP credit assignment. \textbf{Top}: two episodes that differ in restraint under scarcity. In $\omega_1$, agents consume lone apples and incur a lower efficiency score; in $\omega_2$, agents avoid lone-apple consumption and achieve a higher efficiency score, so $\omega_2 \succ \omega_1$. \textbf{Bottom}: the same episode-level preference label drives both MARP variants. In Joint-Episode, the sum of predicted rewards over $\omega_2$ is pushed above that of $\omega_1$, lowering rewards for state--action pairs that consume lone apples. In Local-Trajectory, the same ordering is inherited at the trajectory level, suppressing lone-apple consumption from each agent's perspective.}
  \label{fig:toy-example}
\end{figure}

To illustrate how a single episode-level preference label drives both variants, consider a small example in the Harvest setting (Figure~\ref{fig:toy-example}).

\paragraph{Toy example.} Consider two episodes, $\omega_1$ and $\omega_2$, each with three agents. In $\omega_1$, agents prematurely consume lone apples, leading to lower efficiency, while in $\omega_2$, agents avoid consuming lone apples and achieve higher efficiency. Thus, $\omega_2 \succ \omega_1$.

\paragraph{Joint-Episode.} The reward model $\hat{r}_\theta$ is trained so that
\[
\sum_{(o,a)\in \omega_2} \hat{r}_\theta(o,a) \;>\; \sum_{(o,a)\in \omega_1} \hat{r}_\theta(o,a),
\]
which assigns lower predicted rewards to state-action pairs corresponding to lone-apple consumption.

\paragraph{Local-Trajectory.} Training operates on individual trajectories $\tau^{(i)}_\omega$, but inherits the same ordering. Since trajectories from $\omega_1$ typically include lone-apple consumption while those from $\omega_2$ do not, the model is again pushed to assign lower reward to this behavior.

\subsection{MARP architecture details}

\begin{figure*}[htbp]
  \includegraphics[width=0.7\textwidth]{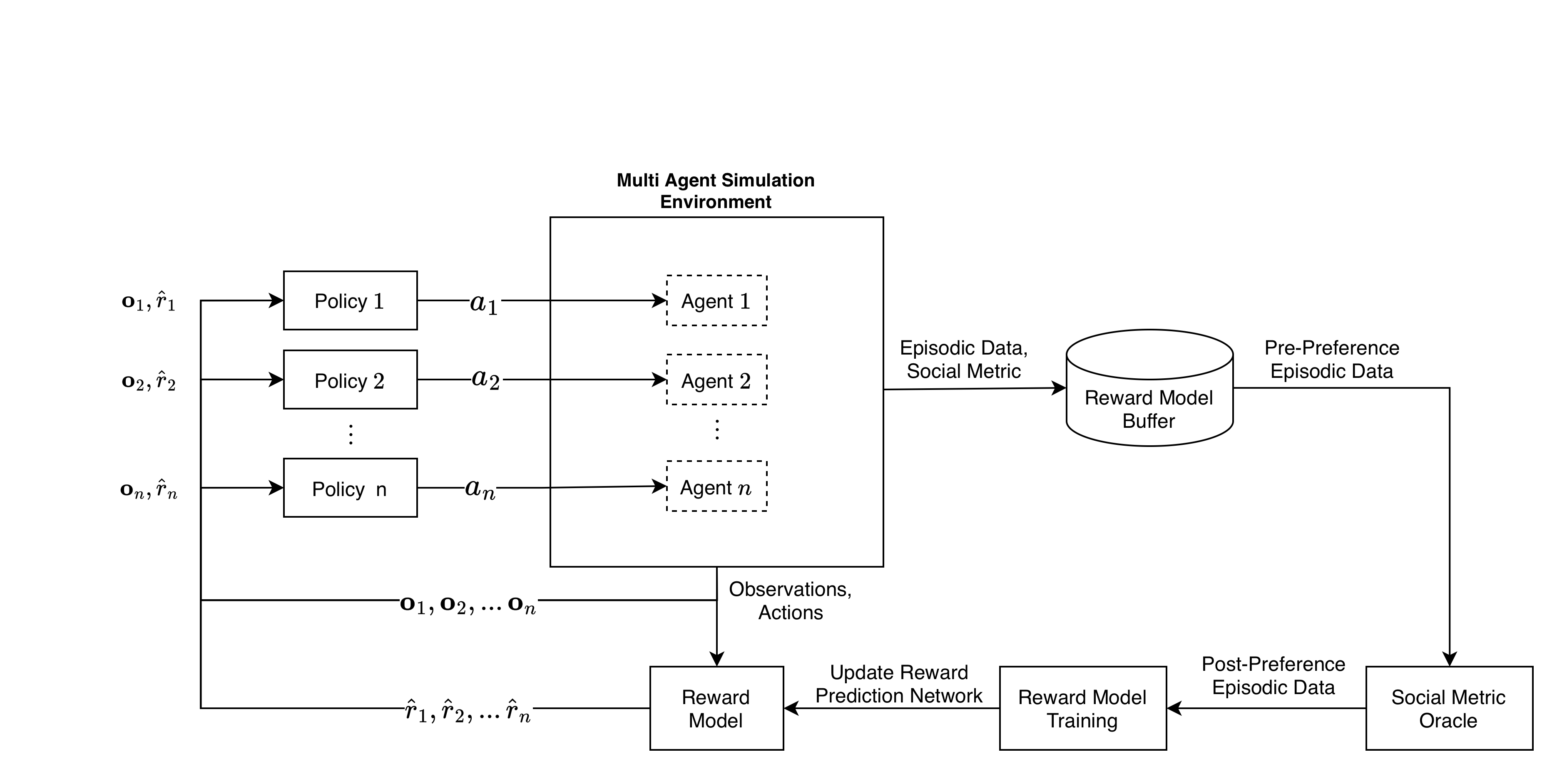}
  \centering
  \caption{Centralized Multi-Agent Reward Prediction (MARP) framework architecture. Multiple agents interact with the simulation environment using decentralized policies, generating episodes stored in a buffer. A social metric oracle evaluates complete episodes to produce preference data, which is used to train a shared reward model. The learned reward predictor then provides per-agent reward signals for policy optimization.}
  \label{fig:RP-Diagram}
\end{figure*}

In this paper, we implement MARP using a centralized reward model together with decentralized agent policies (Figure~\ref{fig:RP-Diagram}). Each agent $i \in \{1,\dots,n\}$ is equipped with its own policy $\pi_i$, while all agents share a single learned reward model $\hat{r}_\theta$ parameterized by $\theta$. At each timestep $t$, agent $i$ selects an action
\[
a_t^{(i)} \sim \pi_i(\cdot \mid o_t^{(i)}),
\]
based solely on its local observation $o_t^{(i)}$. The shared reward model then produces an individual predicted reward
\[
\hat{r}_t^{(i)} = \hat{r}_\theta\!\left(o_t^{(i)}, a_t^{(i)}\right),
\]
which is used as the learning signal for agent $i$.

Training proceeds iteratively in three stages (Figure~\ref{fig:RP-Diagram}). 
Agents first interact with the environment using their decentralized policies 
$\{\pi_i\}_{i=1}^n$, producing an episode 
$\omega = \{\tau^{(1)}, \dots, \tau^{(n)}\}$, which is evaluated by a social metric 
$\phi(\omega)$ and stored in a buffer $\mathcal{B}$. 
Using the current reward model $\hat{r}_\theta$, each policy $\pi_i$ is updated 
independently via RL to maximize the expected cumulative 
predicted reward
\[
\mathbb{E}_{\pi_i}\!\left[\sum_{t=0}^T \hat{r}_t^{(i)}\right].
\]

Next, we update the reward model using the sampled comparisons, where the compared inputs depend on the training strategy. 
In joint-episode, the model is trained on episode-level inputs (an aggregated representation of $\omega$) with labels from $\mathbb{O}_\omega$.
In Local-Trajectory Inference, the model is trained on individual agent trajectories $\tau^{(i)}$ (local-view inputs) while inheriting the episode ordering via $\mathbb{O}_\tau$.
In both cases, comparison weights are scaled by the preference magnitude $\delta_{ij}$ as in Equation~\ref{eq:6}.


The reward model architecture is illustrated in Figure~\ref{fig:RM-architecture}, Appendix~A. We interpret $\hat{r}_\theta: \mathcal{O} \times \mathcal{A} \rightarrow \mathbb{R}$ as a latent factor explaining oracle judgments, where preferences between compared sequences $x, y$ are determined by:
\begin{equation}
x \succ y \iff \sum_{(o,a) \in x} \hat{r}_\theta(o, a) > \sum_{(o,a) \in y} \hat{r}_\theta(o, a)
\end{equation}

Following the Bradley--Terry formulation commonly used in preference-based reward learning \citep{christiano2017deep}, we convert the latent reward sums into a probability of preference via a softmax. Let $s_x = \sum_{(o,a) \in x} \hat{r}_\theta(o, a)$ and $s_y = \sum_{(o,a) \in y} \hat{r}_\theta(o, a)$. Then:

\begin{equation}
P(x \succ y) = \frac{\exp(s_x)}{\exp(s_x) + \exp(s_y)}.
\end{equation}

This probability is what the reward model is trained against (Equation~\ref{eq:6}), so that the model's preference judgments align in expectation with the oracle's.

Our two strategies differ in how sequences are defined. In \textbf{Local-Trajectory} training, the oracle $\mathbb{O}_\tau$ is applied to pairs $(x,y)$ of individual agent trajectories, with $x,y=\tau^{(i)}_{\omega_1},\tau^{(i)}_{\omega_2}$. In contrast, in \textbf{Joint-Episode} training, the oracle $\mathbb{O}_\omega$ is applied to pairs $(x,y)$ of aggregated multi-agent episodes, $x, y = \omega_1,\ \omega_2.$

The reward model is trained using a weighted binary cross-entropy loss over pairwise comparisons $(x,y)$ sampled from the dataset $\mathcal{B}$, with labels $\mu \in \{0,1\}$ indicating whether $x \succ y$. Each comparison is assigned a weight $\delta'_{xy}$, obtained by applying a softmax to standardized preference magnitudes, such that the contribution of each sample in a batch is proportional to the magnitude of the performance gap between the winning and losing outcomes with respect to the social metric.

The two training strategies differ in input format as well as in how preferences are generated: in \textbf{Joint-Episode}, inputs are full episodes and preferences come from the episode-level oracle $\mathbb{O}\omega$; in \textbf{Local-Trajectory Inference}, inputs are individual agent trajectories and preferences are inferred via the trajectory-level oracle $\mathbb{O}_\tau$.

The model parameters $\theta$ are optimized via the following loss, using binary cross entropy (BCE):
\begin{equation} \label{eq:6}
\mathcal{L}(\theta) = \sum_{(x,y, \mu, \delta'_{xy}) \in \mathcal{B}} \delta'_{xy} \cdot \text{BCE}\big(\mu, P(x \succ y)\big)
\end{equation}


We employed a custom implementation of independent PPO (IPPO) across all experiments; agent model parameters appear in Appendix~A, Table~\ref{tab:ippo_hyperparams}, and reward model parameters (both methods) in Appendix~A, Table~\ref{tab:reward_model_hyperparams}.


\section{Results}
\label{sec:Results}

\subsection{From Reward Function to Reward Learning}

\citet{perolat2017multi} demonstrate that when all agents are trained with a simple egocentric reward, receiving 1 for consuming an apple and 0 otherwise, learning dynamics initially lead to a tragedy of the commons, followed by a gradual emergence of avoidance strategies. We successfully reproduced these results using deep Q-Network (DQN; Figure~\ref{fig:dqn-baseline}, Appendix~B). However, DQN requires a substantially larger number of training steps to reach stable behavior. Since contemporary MARL research has largely transitioned to policy-gradient methods, we base our main experimental results on PPO.

When agents are trained directly with PPO using the environment-defined reward, we consistently observe convergence to the tragic equilibrium, characterized by aggressive over-harvesting and an inability to recover cooperative dynamics. This failure mode is robust across hyperparameter settings. In particular, PPO-trained agents fail to transition out of the greedy phase once collective over-exploitation has emerged, even with prolonged training. These observations suggest that, in the Commons Game, standard on-policy PPO is insufficient for resolving the social dilemma when learning is driven solely by environment rewards, whereas off-policy baselines exhibit qualitatively different long-term behavior (also see Discussion).

In contrast, when using MARP, the  Joint-Episode and Local-Trajectory Inference consistently enable agents to escape the tragic equilibrium and transition toward stable cooperative regimes (see Figure~\ref{fig:MARP-efficiency-all-metrics} below; videos of late-episode behavior under the Efficiency-only objective are available for the Joint-Episode\footnote{\url{https://youtu.be/2fcIcd5sPOk}} and Local-Trajectory\footnote{\url{https://youtu.be/DLp-U1RmBOs}} variants). 

\begin{figure}[htbp]
  \includegraphics[width=\columnwidth]{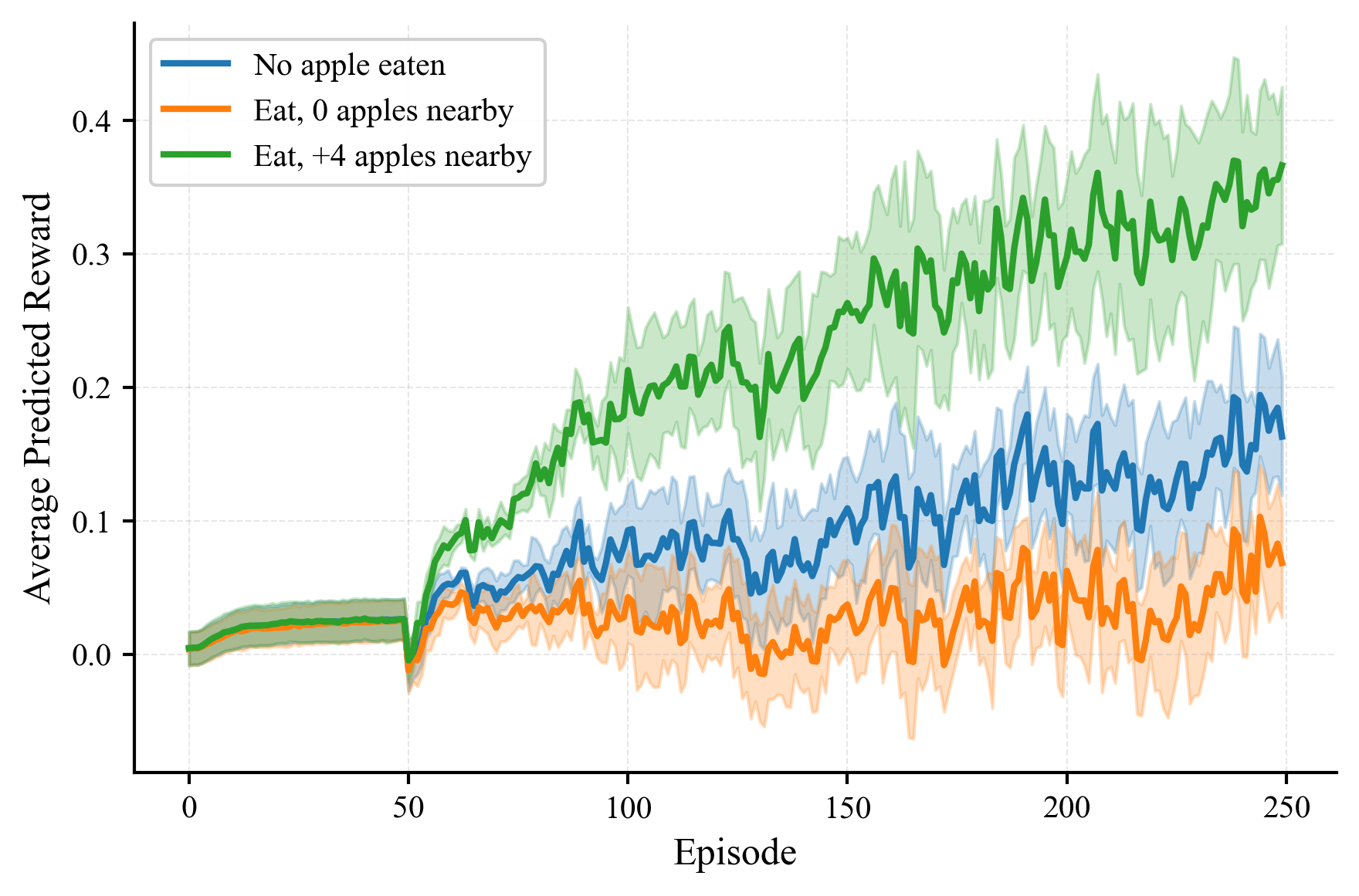}
  \centering
  \caption{Reward prediction using the Joint-Episode method. The figure compares three conditions: (i) no apple is consumed, (ii) an apple is consumed under local scarcity, with no neighboring apples present, and (iii) an apple is consumed when all four surrounding apples are present. Rewards are averaged across the five agents, and error bars indicate standard error.}

  \label{fig:apple-num}
\end{figure}

Importantly, we observe that the Joint-Episode method is able to learn a nuanced reward model that captures fine-grained environmental structure in a purely data-driven manner -- the learned reward assigns lower value to consuming apples under local scarcity, effectively discouraging harvesting when the surrounding apple density is low (Figure~\ref{fig:apple-num}).

We assessed differences in reward predictions using a two-way repeated-measures ANOVA over condition type and episode group, averaging rewards across agents. The analysis revealed significant main effects of condition type and episode group, as well as a strong interaction between them ($F>15$, $p<0.001$ in all cases), with all effects remaining significant under Greenhouse–Geisser correction, indicating that reward predictions differ across conditions and evolve over training. For the Local-Trajectory method, only the effect of episode group was significant, with no condition-type effect or interaction (Figure~\ref{fig:lt-reward}, Appendix~B).

\subsection{Learning to Align with Multiple and Composite Social Metrics}

\begin{figure}[htbp]
  \includegraphics[width=\columnwidth]{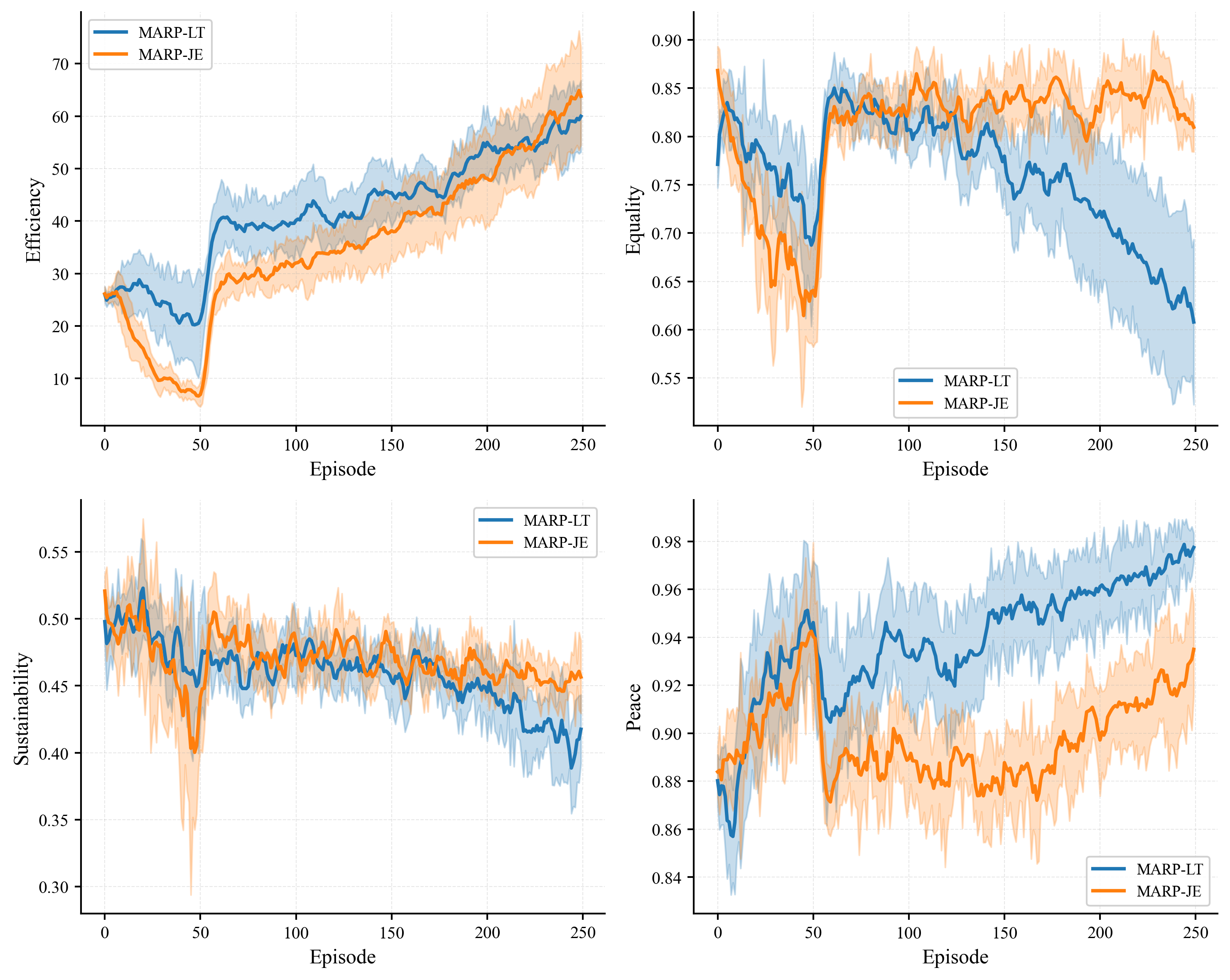}
  \centering
  \caption{Social metrics under Efficiency-only optimization for the two MARP variants. Error bars denote standard error over five runs.}
  \label{fig:MARP-efficiency-all-metrics}
\end{figure}

So far, we have optimized for efficiency. Because MARP decouples reward learning from the optimization objective, the same framework can be applied to alternative or composite social metrics without modifying the learning algorithm. Figure~\ref{fig:MARP-efficiency-all-metrics} reports all social metrics under efficiency optimization and shows no consistent advantage of one method over the other. 

\begin{figure}[htbp]
  \includegraphics[width=\columnwidth]{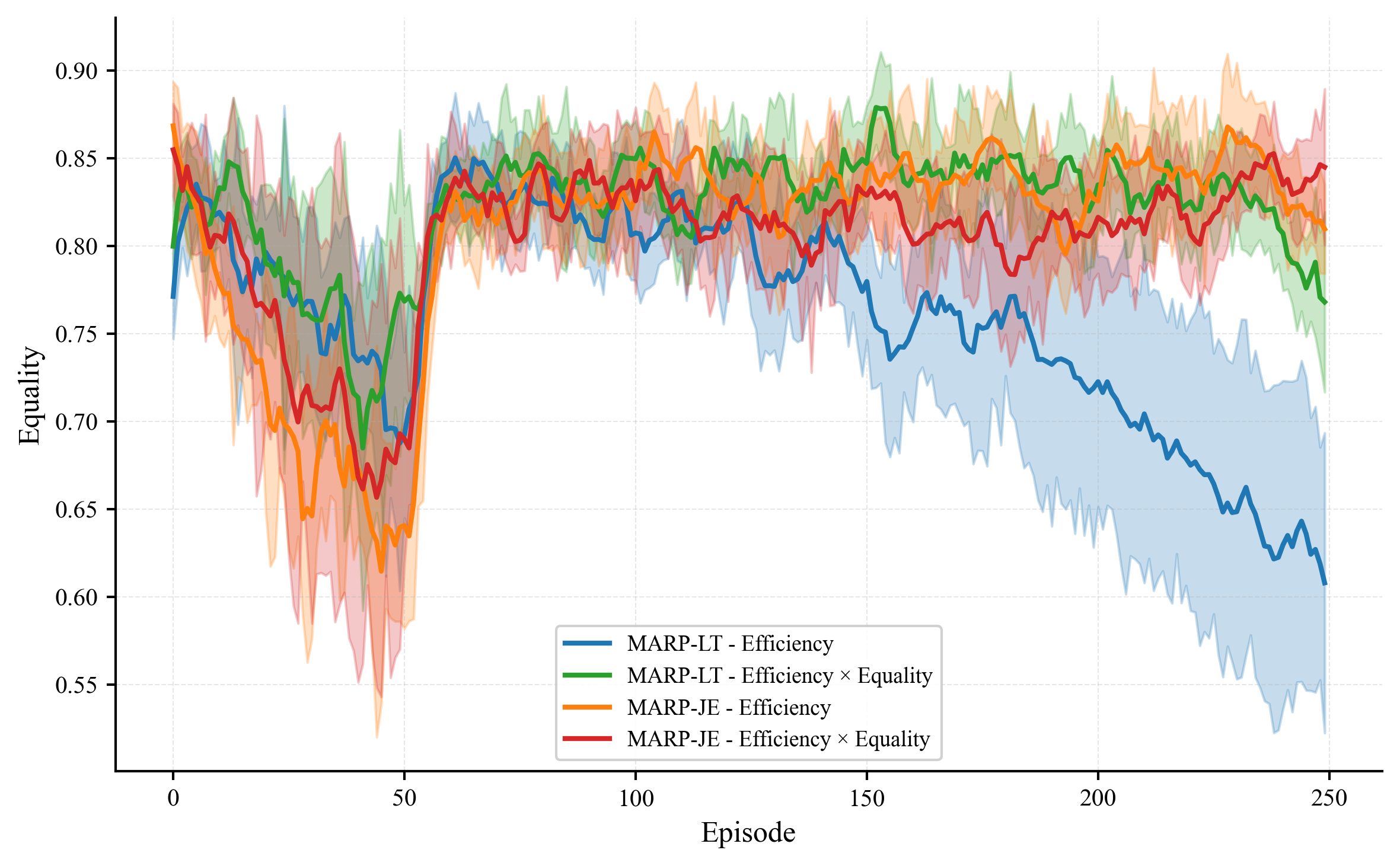}
  \centering
  \caption{Equality under MARP for efficiency-only and efficiency$\times$equality objectives. Error bars show standard error over five runs.}
\label{fig:E-Eq-metric-Eq}
\end{figure}

Under the Local-Trajectory variant, Efficiency is achieved but Equality steadily degrades. Our framework allows optimizing for $\text{Efficiency} \times \text{Equality}$ to satisfy multiple social goals simultaneously. Figure~\ref{fig:E-Eq-metric-Eq} reports results for the Equality metric. Although runs optimizing the combined objective initially descend further into the tragic regime, both variants ultimately maintain high levels of Equality; despite observing one agent at a time, the reward model learns to capture a group-level notion of equality.

Across all combinations of dual social objectives, we observe no consistent advantage of either Local-Trajectory Inference or Joint-Episode over the other. More importantly, optimizing for a secondary social metric consistently succeeds: for all methods and all metric pairs, the dual-objective setting yields a higher trend for the secondary metric, without degrading efficiency. Videos of late-episode behavior under the Efficiency-Peace objective are available for the Joint-Episode\footnote{\url{https://youtu.be/69I0zKO9e34}} and Local-Trajectory\footnote{\url{https://youtu.be/Nf8onEPsp6U}} variants. Detailed results for all social metrics and objective combinations are provided in Appendix~B (Figures~\ref{fig:social-metrics-all} and~\ref{fig:nv-ia-E-vs-dual}).

\section{Discussion}

Our experiments suggest that Multi-Agent Reward Prediction (MARP) offers a proof-of-concept mechanism for regulating emergent behavior in decentralized multi-agent systems. Across both Joint-Episode and Local-Trajectory inference strategies, MARP outperforms PPO trained directly on the environment reward in the Commons Game, demonstrating that preference-based reward modeling can reliably steer behavior toward socially desirable regimes. More broadly, MARP is a methodological contribution rather than one tied to a specific strategy or domain: by changing only the social metric used by the preference oracle, the same framework, architecture, and training procedure can regulate diverse collective objectives, including composite goals, without modifying agent policies or environment dynamics. The reward model itself is retrained for each target metric. While the method is designed to be applicable across multi-agent settings, the present empirical evaluation is limited to a single environment, and we therefore present MARP as a proof of concept rather than a fully validated general-purpose alignment mechanism.

When trained directly with PPO on the environment reward, agents converged to the tragic equilibrium and did not recover cooperation. This may reflect a limitation of on-policy optimization in exploration-constrained social dilemmas, where the data distribution becomes self-reinforcing once greedy behavior dominates. Consistent with this view, prior work shows that adding experience replay to PPO can improve performance in sparse-reward settings~\citep{wang2024replayppo}. In contrast, MARP’s episodic buffer for reward modeling may preserve information from pre-collapse regimes, sustaining gradients toward cooperative behavior, similar to how preference-based reward shaping has improved exploration in classic RL settings~\citep{christiano2017deep}.



A central question is whether the learned reward model faithfully reflects the preference oracle or induces unintended behavioral drift. We evaluate this at both behavioral and representational levels. Behaviorally, MARP-trained agents consistently move social metrics in the intended direction across Efficiency, Peace, Sustainability, and Equality. Representationally, the learned reward function recovers the oracle’s ordinal distinctions, for example favoring apples with high regeneration potential and assigning higher value to non-violent actions under combined objectives. 


\subsection{Limitations}

MARP relies on episodic, metric-based supervision, which introduces two structural limitations. First, episode-level evaluation reduces temporal resolution, making it difficult to assign credit to short or rare actions with disproportionate social impact. Second, alignment depends on predefined social metrics, which may be insufficient in domains with ambiguous, evolving, or contested objectives, or where trade-offs cannot be captured by a fixed scalar measure.

In our experiments, these limitations are mitigated by using deterministic, metric-based preference oracles over complete episodes, providing a controlled setting for isolating the technical challenge of translating global evaluations into local reward signals. However, this substantially simplifies the alignment problem relative to real-world settings, where human feedback is noisy, context-dependent, and inconsistent. MARP should therefore be viewed as an intermediate step toward multi-agent alignment rather than a complete solution.

Beyond these structural limitations, three further caveats apply to our present results. First, the credit-assignment behavior of MARP is established empirically rather than derived from theoretical guarantees: optimizing the learned reward consistently shifts behavior toward the target metrics, but we do not claim that the model recovers a correct or uniquely identifiable per-step decomposition of the global preference. Second, Local-Trajectory inference inherits a single episode-level label across all agent trajectories within an episode, which provides a coarse and potentially noisy training signal. In our setting, agents share a reward model and tend to learn relatively homogeneous policies, which reduces but does not eliminate the ambiguity in attributing outcomes to specific agents. Third, the method's apparent effectiveness may depend on favorable properties of the Harvest setting --- in particular, long-horizon consequences of local actions, homogeneous agents, and deterministic metric-based oracles. Generalization to settings with heterogeneous agents, finer-grained temporal credit-assignment needs, or stochastic human feedback remains an open question.

\subsection{Ethical considerations}

Social simulations are widely used to study human behavior, but extending them toward regulation of emergent behavior introduces both significant opportunities and risks. In principle, frameworks such as MARP could be applied to influence human behavior in large-scale digital environments, a concern that is particularly salient for social media platforms, where algorithmic systems already shape behavior by optimizing engagement and attention at scale~\citep{tufekci2015algorithmic, bakshy2015exposure, zuboff2023age}.


Our goal is not to advocate behavioral control, but to highlight that existing systems already regulate collective behavior through narrow objectives such as engagement maximization. In contrast, MARP enables optimization over a broader class of explicit and interpretable objectives, including social welfare, fairness, or sustainability. At the same time, the ability to steer collective dynamics using learned reward models raises serious ethical and legal concerns. Without appropriate safeguards, transparency, and oversight, such methods could be misused to manipulate populations or advance harmful agendas, underscoring the need to situate technical advances in multi-agent alignment within a broader ethical and regulatory context.

\subsection{Future Work}

Future work should extend MARP to more realistic alignment settings. A natural direction is to decentralize reward modeling, allowing agents to maintain local reward predictors tailored to their observations or roles, which may improve robustness, interpretability, and scalability. Closely related is evaluating MARP with heterogeneous agents, where differences in capabilities or roles challenge shared reward assumptions and test how alignment generalizes across diverse populations.

Beyond the Commons Game, MARP should be evaluated on a broader range of multi-agent benchmarks, particularly those with partial observability, conflicting incentives, or emergent role differentiation, to assess generalization beyond a single social dilemma.

Finally, framing alignment as reward prediction over joint behaviors and long-horizon outcomes opens new avenues for preference-based fine-tuning in decentralized systems, while underscoring the need for careful governance as such methods move beyond controlled simulations.





\bibliographystyle{ijcai25}
\bibliography{reference}

\begin{thebibliography}{}

\bibitem[\protect\citeauthoryear{Axelrod}{1997}]{axelrod1997advancing}
Robert Axelrod.
\newblock Advancing the art of simulation in the social sciences.
\newblock In {\em Simulating social phenomena}, pages 21--40. Springer, 1997.

\bibitem[\protect\citeauthoryear{Axtell and Farmer}{2025}]{axtell2025agent}
Robert~L Axtell and J~Doyne Farmer.
\newblock Agent-based modeling in economics and finance: Past, present, and
  future.
\newblock {\em Journal of Economic Literature}, 63(1):197--287, 2025.

\bibitem[\protect\citeauthoryear{Bakshy \bgroup \em et al.\egroup
  }{2015}]{bakshy2015exposure}
Eytan Bakshy, Solomon Messing, and Lada~A Adamic.
\newblock Exposure to ideologically diverse news and opinion on facebook.
\newblock {\em Science}, 348(6239):1130--1132, 2015.

\bibitem[\protect\citeauthoryear{Christiano \bgroup \em et al.\egroup
  }{2017}]{christiano2017deep}
Paul~F Christiano, Jan Leike, Tom Brown, Miljan Martic, Shane Legg, and Dario
  Amodei.
\newblock Deep reinforcement learning from human preferences.
\newblock {\em Advances in neural information processing systems}, 30, 2017.

\bibitem[\protect\citeauthoryear{Gronauer and
  Diepold}{2022}]{gronauer2022multi}
Sven Gronauer and Klaus Diepold.
\newblock Multi-agent deep reinforcement learning: a survey.
\newblock {\em Artificial Intelligence Review}, 55(2):895--943, 2022.

\bibitem[\protect\citeauthoryear{Hughes \bgroup \em et al.\egroup
  }{2018}]{hughes2018inequity}
Edward Hughes, Joel~Z. Leibo, Matthew~G. Phillips, Karl Tuyls, Edgar~A.
  Du{\'e}{\~n}ez-Guzm{\'a}n, Antonio~Garc{\'i}a Casta{\~n}eda, Iain Dunning,
  Tina Zhu, Kevin~R. McKee, Raphael Koster, Heather Roff, and Thore Graepel.
\newblock Inequity aversion improves cooperation in intertemporal social
  dilemmas, 2018.

\bibitem[\protect\citeauthoryear{Jaques \bgroup \em et al.\egroup
  }{2019}]{jaques2019social}
Natasha Jaques, Angeliki Lazaridou, Edward Hughes, Caglar Gulcehre, Pedro
  Ortega, DJ~Strouse, Joel~Z Leibo, and Nando De~Freitas.
\newblock Social influence as intrinsic motivation for multi-agent deep
  reinforcement learning.
\newblock In {\em International conference on machine learning}, pages
  3040--3049. PMLR, 2019.

\bibitem[\protect\citeauthoryear{Kerr \bgroup \em et al.\egroup
  }{2021}]{kerr2021covasim}
Cliff~C Kerr, Robyn~M Stuart, Dina Mistry, Romesh~G Abeysuriya, Katherine
  Rosenfeld, Gregory~R Hart, Rafael~C N{\'u}{\~n}ez, Jamie~A Cohen, Prashanth
  Selvaraj, Brittany Hagedorn, et~al.
\newblock Covasim: an agent-based model of covid-19 dynamics and interventions.
\newblock {\em PLOS Computational Biology}, 17(7):e1009149, 2021.

\bibitem[\protect\citeauthoryear{Leibo \bgroup \em et al.\egroup
  }{2017}]{leibo2017multi}
Joel~Z Leibo, Vinicius Zambaldi, Marc Lanctot, Janusz Marecki, and Thore
  Graepel.
\newblock Multi-agent reinforcement learning in sequential social dilemmas.
\newblock {\em arXiv preprint arXiv:1702.03037}, 2017.

\bibitem[\protect\citeauthoryear{Leike \bgroup \em et al.\egroup
  }{2018}]{leike2018scalable}
Jan Leike, David Krueger, Tom Everitt, Miljan Martic, Vishal Maini, and Shane
  Legg.
\newblock Scalable agent alignment via reward modeling: a research direction.
\newblock {\em arXiv preprint arXiv:1811.07871}, 2018.

\bibitem[\protect\citeauthoryear{Ngo \bgroup \em et al.\egroup
  }{2022}]{ngo2022alignment}
Richard Ngo, Lawrence Chan, and S{\"o}ren Mindermann.
\newblock The alignment problem from a deep learning perspective.
\newblock {\em arXiv preprint arXiv:2209.00626}, 2022.

\bibitem[\protect\citeauthoryear{Ouyang \bgroup \em et al.\egroup
  }{2022}]{ouyang2022training}
Long Ouyang, Jeffrey Wu, Xu~Jiang, Diogo Almeida, Carroll Wainwright, Pamela
  Mishkin, Chong Zhang, Sandhini Agarwal, Katarina Slama, Alex Ray, et~al.
\newblock Training language models to follow instructions with human feedback.
\newblock {\em Advances in neural information processing systems},
  35:27730--27744, 2022.

\bibitem[\protect\citeauthoryear{Perolat \bgroup \em et al.\egroup
  }{2017}]{perolat2017multi}
Julien Perolat, Joel~Z Leibo, Vinicius Zambaldi, Charles Beattie, Karl Tuyls,
  and Thore Graepel.
\newblock A multi-agent reinforcement learning model of common-pool resource
  appropriation.
\newblock {\em Advances in neural information processing systems}, 30, 2017.

\bibitem[\protect\citeauthoryear{Stiennon \bgroup \em et al.\egroup
  }{2020}]{stiennon2020learning}
Nisan Stiennon, Long Ouyang, Jeffrey Wu, Daniel Ziegler, Ryan Lowe, Chelsea
  Voss, Alec Radford, Dario Amodei, and Paul~F Christiano.
\newblock Learning to summarize with human feedback.
\newblock {\em Advances in Neural Information Processing Systems},
  33:3008--3021, 2020.

\bibitem[\protect\citeauthoryear{Trott \bgroup \em et al.\egroup
  }{2021}]{trott2021building}
Alexander Trott, Sunil Srinivasa, Douwe van~der Wal, Sebastien Haneuse, and
  Stephan Zheng.
\newblock Building a foundation for data-driven, interpretable, and robust
  policy design using the {AI} {E}conomist.
\newblock {\em arXiv preprint arXiv:2108.02904}, 2021.

\bibitem[\protect\citeauthoryear{Tufekci}{2015}]{tufekci2015algorithmic}
Zeynep Tufekci.
\newblock Algorithmic harms beyond facebook and google: Emergent challenges of
  computational agency.
\newblock {\em Colorado Technology Law Journal}, 13:203--218, 2015.

\bibitem[\protect\citeauthoryear{Vinitsky \bgroup \em et al.\egroup
  }{2019}]{SSDOpenSource}
Eugene Vinitsky, Natasha Jaques, Joel Leibo, Antonio Castenada, and Edward
  Hughes.
\newblock An open source implementation of sequential social dilemma games.
\newblock
  \url{https://github.com/eugenevinitsky/sequential_social_dilemma_games/issues/182},
  2019.
\newblock GitHub repository.

\bibitem[\protect\citeauthoryear{Wang \bgroup \em et al.\egroup
  }{2024}]{wang2024replayppo}
Yifan Wang, Yuchen Zhang, Ziming Xu, and Shuo Chen.
\newblock On the importance of replay for policy optimization in sparse-reward
  and discrete action spaces.
\newblock {\em arXiv preprint arXiv:2405.16383}, 2024.

\bibitem[\protect\citeauthoryear{Zheng \bgroup \em et al.\egroup
  }{2022}]{zheng2022aieconomist}
Stephan Zheng, Alexander Trott, Sunil Srinivasa, David~C. Parkes, and Richard
  Socher.
\newblock The {AI} {E}conomist: Taxation policy design via two-level deep
  multiagent reinforcement learning.
\newblock {\em Science Advances}, 8(18):eabk2607, 2022.

\bibitem[\protect\citeauthoryear{Zhu \bgroup \em et al.\egroup
  }{2024}]{zhu2024decoding}
Tianchen Zhu, Yue Qiu, Haoyi Zhou, and Jianxin Li.
\newblock Decoding global preferences: Temporal and cooperative dependency
  modeling in multi-agent preference-based reinforcement learning.
\newblock In {\em Proceedings of the AAAI Conference on Artificial
  Intelligence}, volume~38, pages 17202--17210, 2024.

\bibitem[\protect\citeauthoryear{Zuboff}{2023}]{zuboff2023age}
Shoshana Zuboff.
\newblock The age of surveillance capitalism.
\newblock In {\em Social theory re-wired}, pages 203--213. Routledge, 2023.

\end{thebibliography}

\clearpage
\appendix

\section*{Appendix A: Experimental Hyper-parameters}
\label{app:hyperparams}

\begin{figure*}[!tbp]
  \centering
  \includegraphics[width=0.7\textwidth]{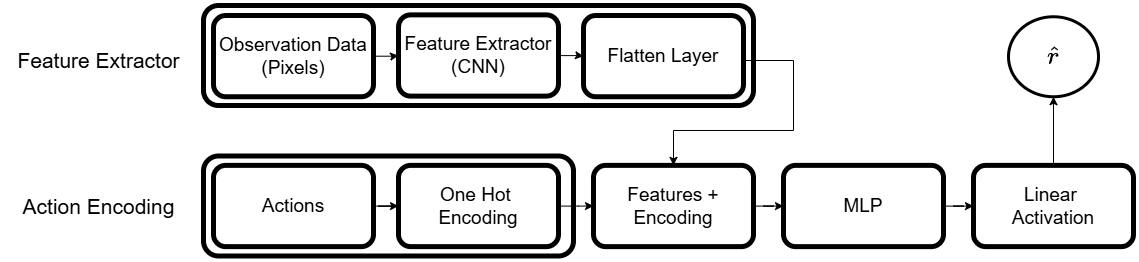}
  \caption{Reward model architecture. Observations are processed through a CNN-based feature extractor and flattened, while actions are mapped to one-hot encoding. The extracted features and action encoding are concatenated and passed through an MLP and a linear layer to compute the predicted reward $\hat{r}$.}
  \label{fig:RM-architecture}
\end{figure*}

This appendix lists the architecture and training hyper-parameters used in all experiments. Figure~\ref{fig:RM-architecture} illustrates the reward model architecture. Table~\ref{tab:ippo_hyperparams} reports the IPPO agent hyper-parameters used by every policy, and Table~\ref{tab:reward_model_hyperparams} reports the hyper-parameters used to train the shared reward model. The same agent and reward-model settings are used across all social-metric objectives reported in the main paper.

\begin{table*}[!tbp]
\centering
\small
\caption{Independent PPO (IPPO) Agent Hyper-parameters.}
\begin{tabular}{lll}
\toprule
\textbf{Parameter} & \textbf{Value} & \textbf{Description} \\
\midrule
Learner & IPPO (PPO) & Independent Proximal Policy Optimization \\
Learning rate & 0.0003 & Optimizer learning rate \\
Batch size & 128 & Number of samples per training batch \\
Discount factor ($\gamma$) & 0.99 & Future reward discounting \\
GAE $\lambda$ & 0.95 & Bias--variance tradeoff for advantage estimation \\
Clip range & 0.2 & PPO policy clipping parameter \\
Entropy coefficient (start) & 0.1 & Initial entropy regularization weight \\
Entropy coefficient (end) & 0.01 & Final entropy regularization weight \\
Value function coefficient & 0.5 & Weight of value loss in PPO objective \\
Value function clip & 10.0 & Clipping range for value updates \\
Rollout steps ($n_{\text{steps}}$) & 512 & Environment steps per policy update \\
Update epochs & 2 & Gradient passes per PPO update \\
Max gradient norm & 0.5 & Gradient clipping threshold \\
Hidden layer size & 256 & Policy and value network hidden dimension \\
Policy & MLP & Multi-layer perceptron policy \\
\bottomrule
\end{tabular}
\label{tab:ippo_hyperparams}
\end{table*}

\begin{table*}[!tbp]
\centering
\small
\caption{Reward Model Training Hyper-parameters.}
\begin{tabular}{lll}
\toprule
\textbf{Parameter} & \textbf{Value} & \textbf{Description} \\
\midrule
Learning rate & 0.0001 & Optimizer learning rate for reward model \\
Batch pairs & 64 & Number of trajectory pairs sampled per update \\
Train steps per update & 50 & Gradient steps per reward model update \\
Update frequency & 1000 & Environment steps between reward model updates \\
Warmup episodes & 50 & Episodes collected before reward model training begins \\
Max episodes in buffer & 5000 & Capacity of preference replay buffer \\
Fully connected layers & $[128 + |\mathcal{A}|,\ 128,\ 1]$ & Reward network architecture \\
\bottomrule
\end{tabular}
\label{tab:reward_model_hyperparams}
\end{table*}

\FloatBarrier
\clearpage
\section*{Appendix B: Additional Results}
\label{app:B}

\begin{figure}[!tb]
  \centering
  \includegraphics[width=\columnwidth]{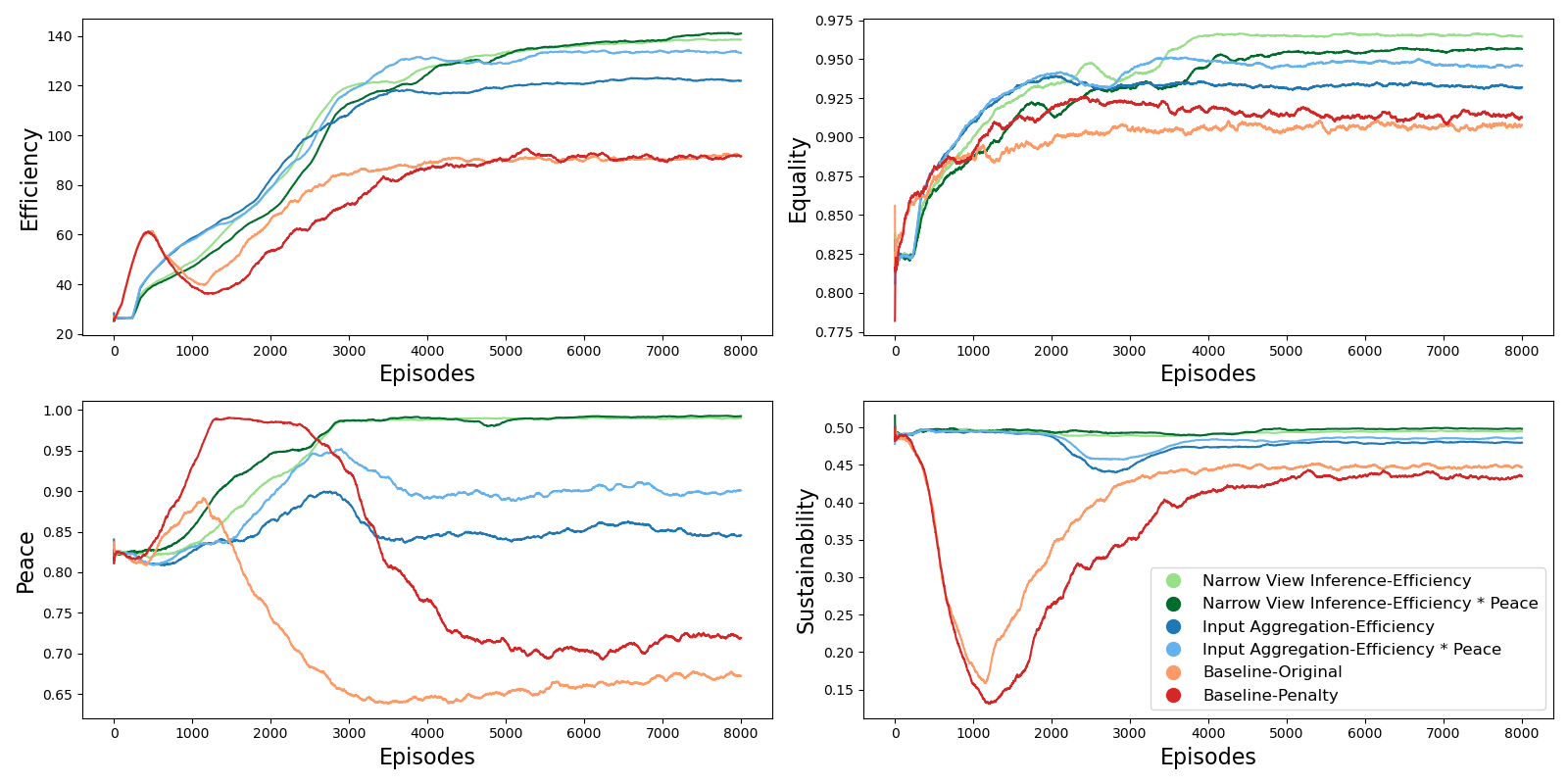}
  \caption{Comparison between DQN with environment-defined rewards, reproducing the setup of \citet{perolat2017multi}, and our reward prediction approach. Results are shown for a single run per method.}
  \label{fig:dqn-baseline}
\end{figure}

\begin{figure}[!tb]
  \centering
  \includegraphics[width=\columnwidth]{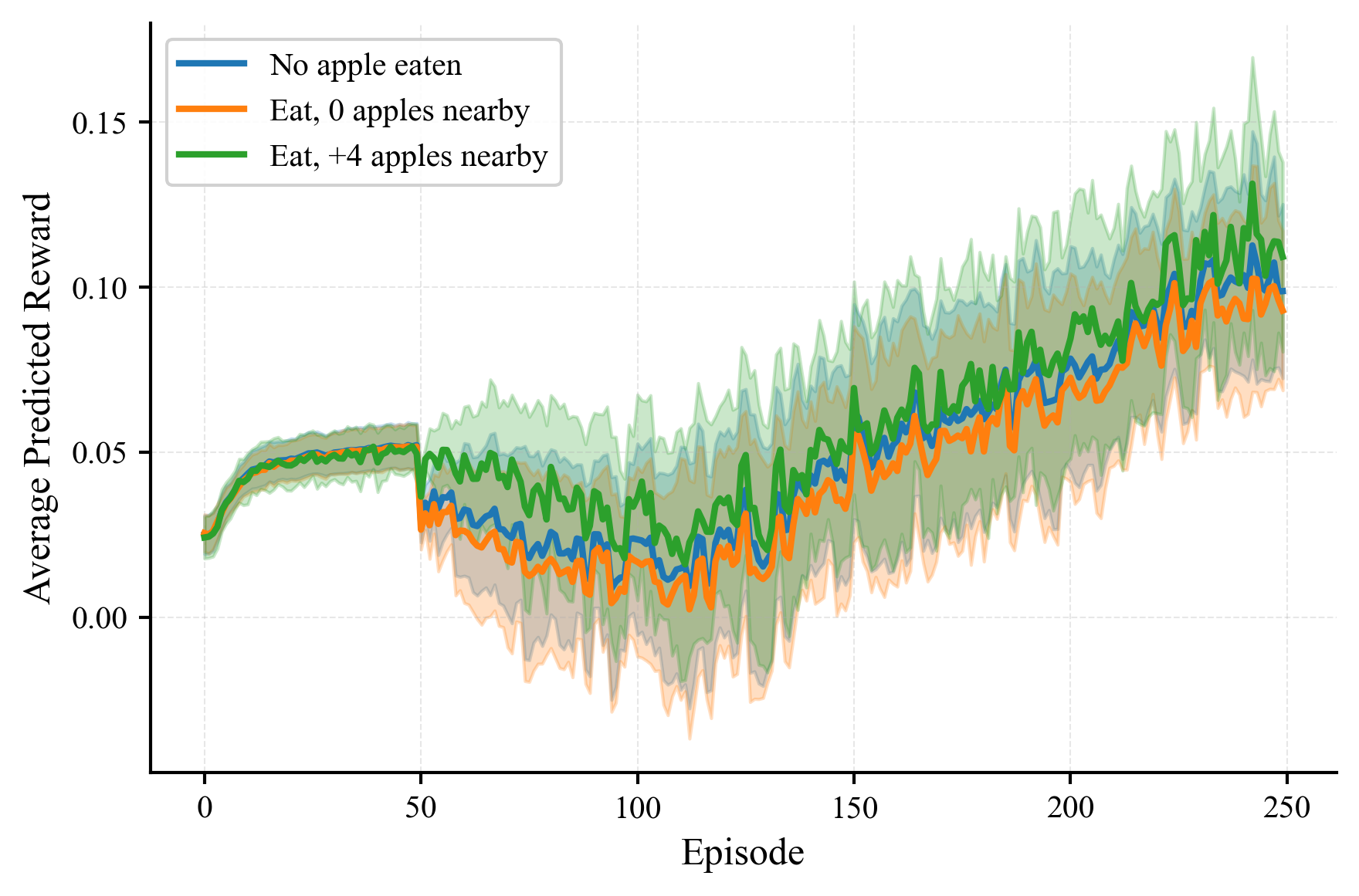}
  \caption{Reward prediction using the Local-Trajectory method. The figure compares three conditions: i) no apple eaten, ii) apple eaten and no other apples in observation window, and iii) apple eaten and there are at least four apples in the observation area. The reward is averaged over the five agents. Error bars indicate standard error.}
  \label{fig:lt-reward}
\end{figure}

\begin{figure*}[!tbp]
    \centering
    \begin{subfigure}[t]{0.45\textwidth}
        \centering
        \includegraphics[width=\linewidth]{social_metrics_gallery_efficiency.png}
        \caption{Efficiency}
        \label{fig:sm-a}
    \end{subfigure}
    \hfill
    \begin{subfigure}[t]{0.45\textwidth}
        \centering
        \includegraphics[width=\linewidth]{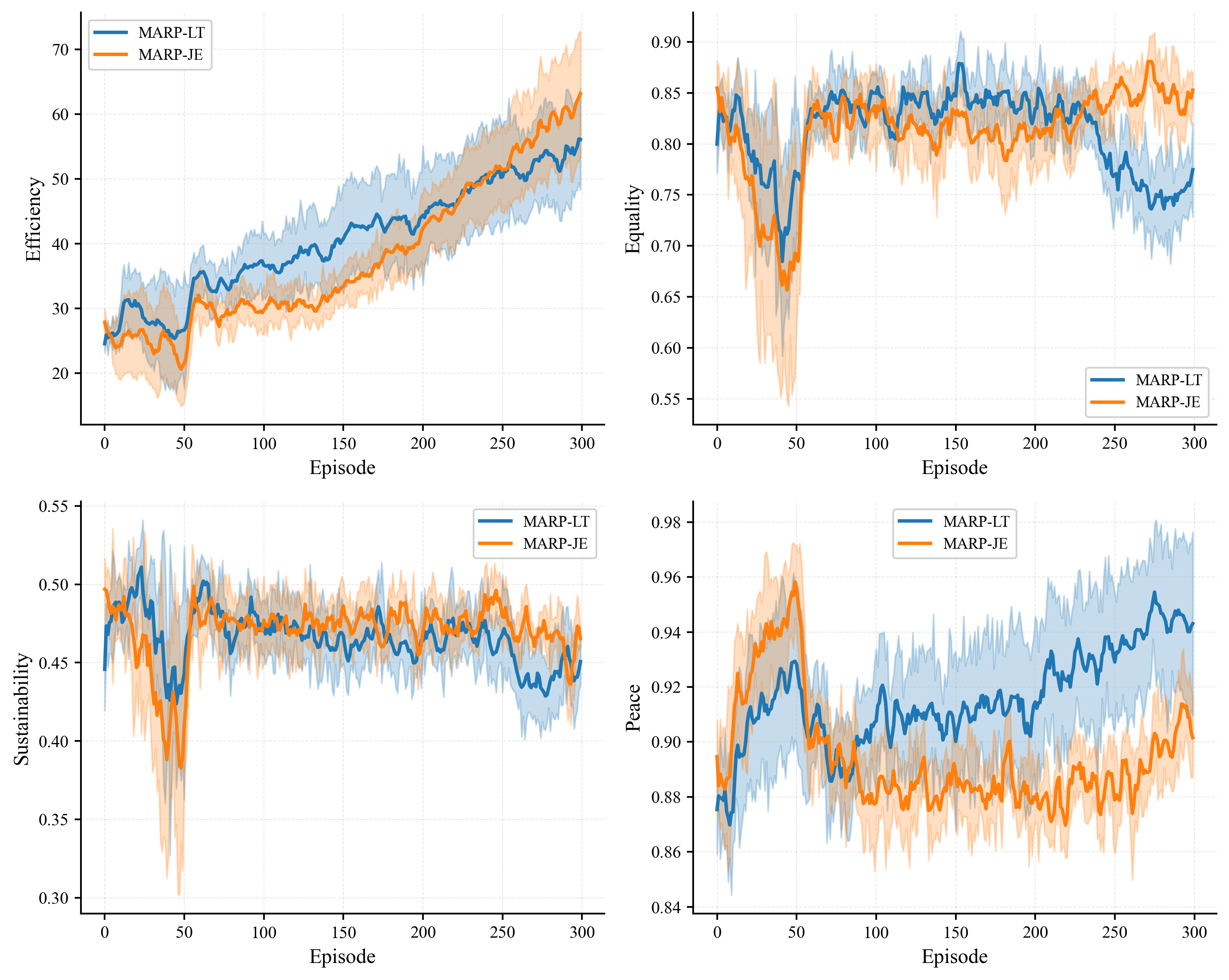}
        \caption{Efficiency $\times$ Equality}
        \label{fig:sm-b}
    \end{subfigure}

    \vspace{0.8em}

    \begin{subfigure}[t]{0.45\textwidth}
        \centering
        \includegraphics[width=\linewidth]{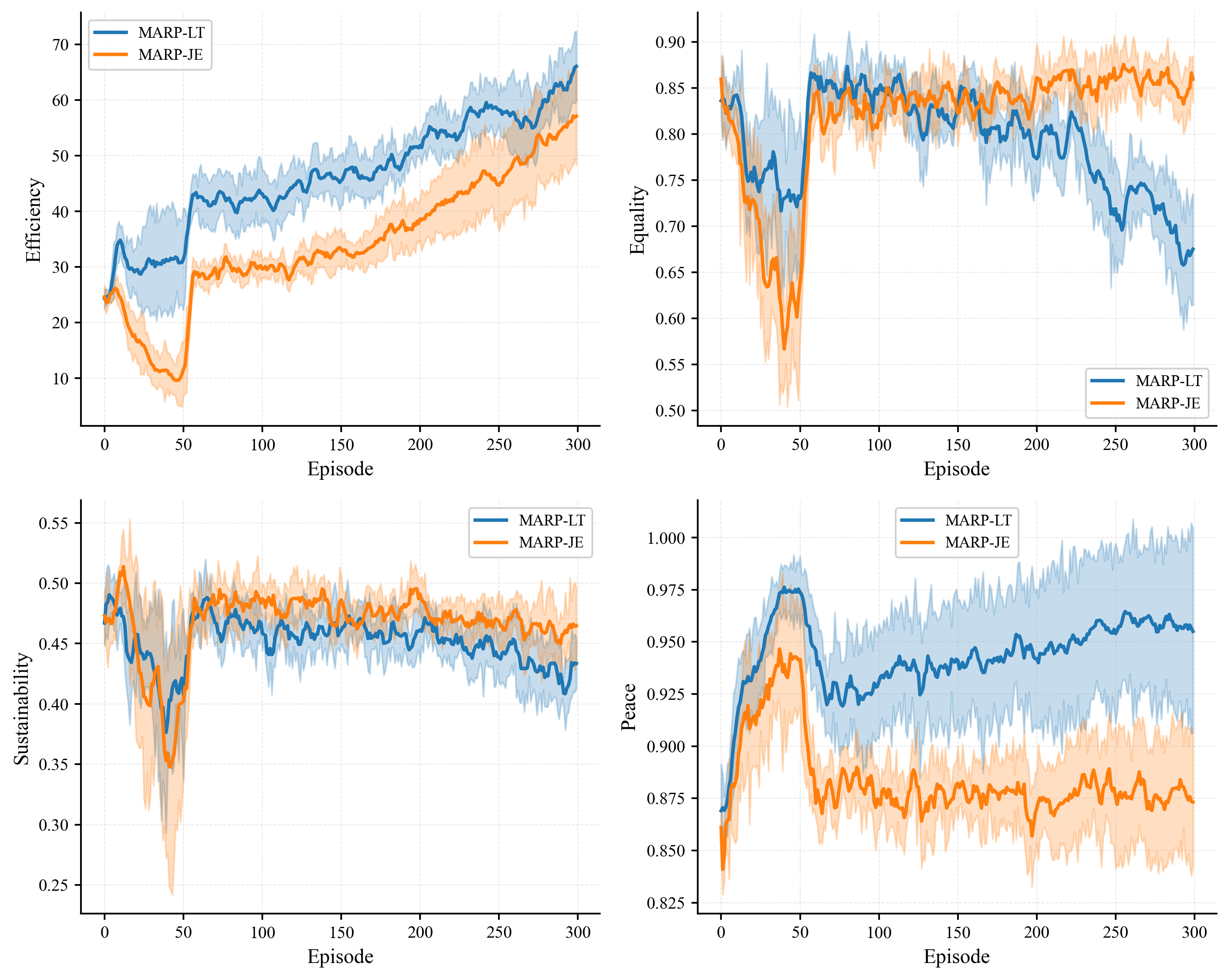}
        \caption{Efficiency $\times$ Sustainability}
        \label{fig:sm-c}
    \end{subfigure}
    \hfill
    \begin{subfigure}[t]{0.45\textwidth}
        \centering
        \includegraphics[width=\linewidth]{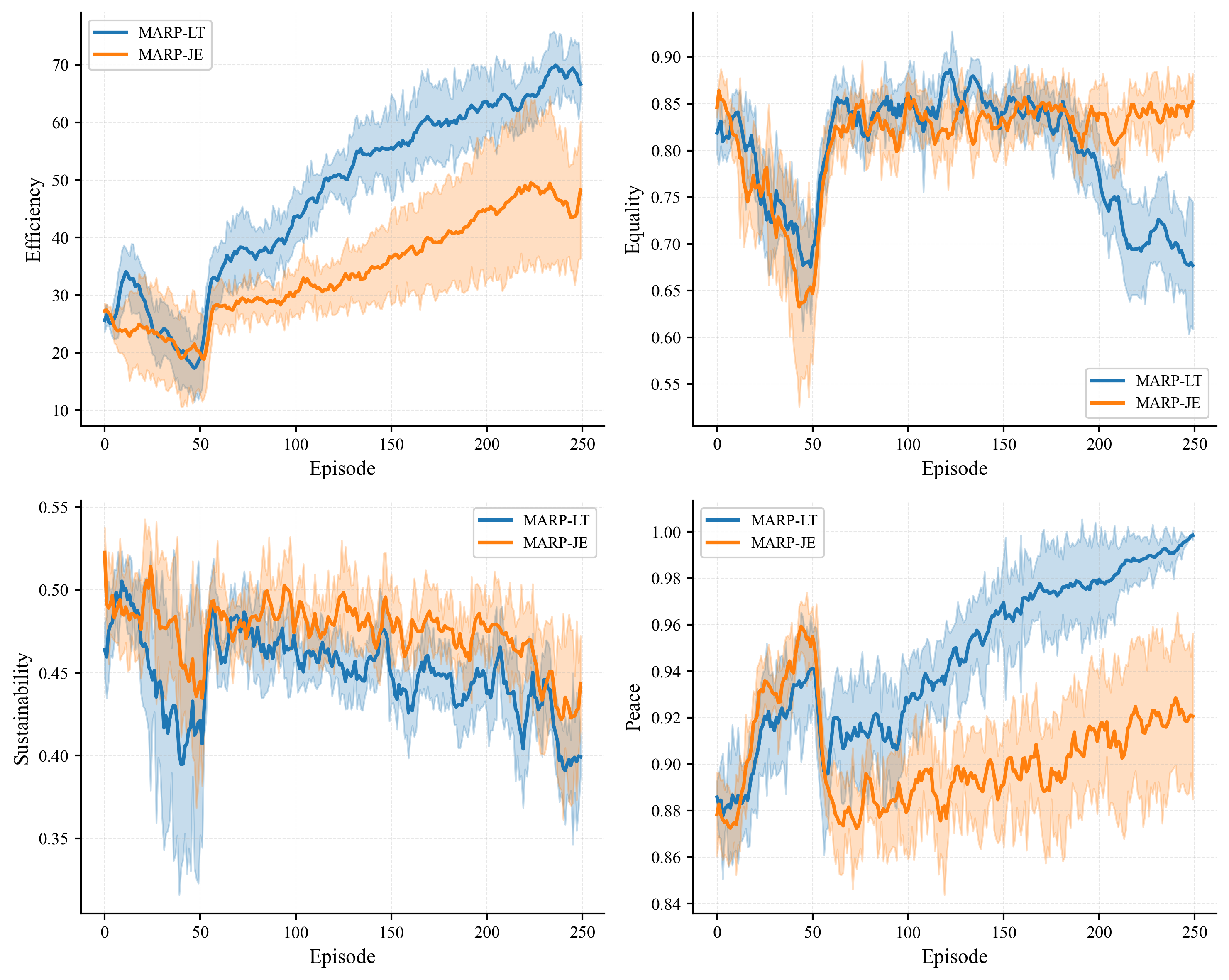}
        \caption{Efficiency $\times$ Peace}
        \label{fig:sm-d}
    \end{subfigure}

    \caption{For each learning objective, we report all social metrics, comparing the two MARP methods. Results are based on five runs; error bars indicate standard error.}
    \label{fig:social-metrics-all}
\end{figure*}

\begin{figure*}[!tbp]
    \centering
    \begin{subfigure}[t]{0.45\textwidth}
        \centering
        \includegraphics[width=\linewidth]{phi_comparison_efficiency_vs_equality.png}
        \caption{Equality}
        \label{fig:phi-a}
    \end{subfigure}
    \hfill
    \begin{subfigure}[t]{0.45\textwidth}
        \centering
        \includegraphics[width=\linewidth]{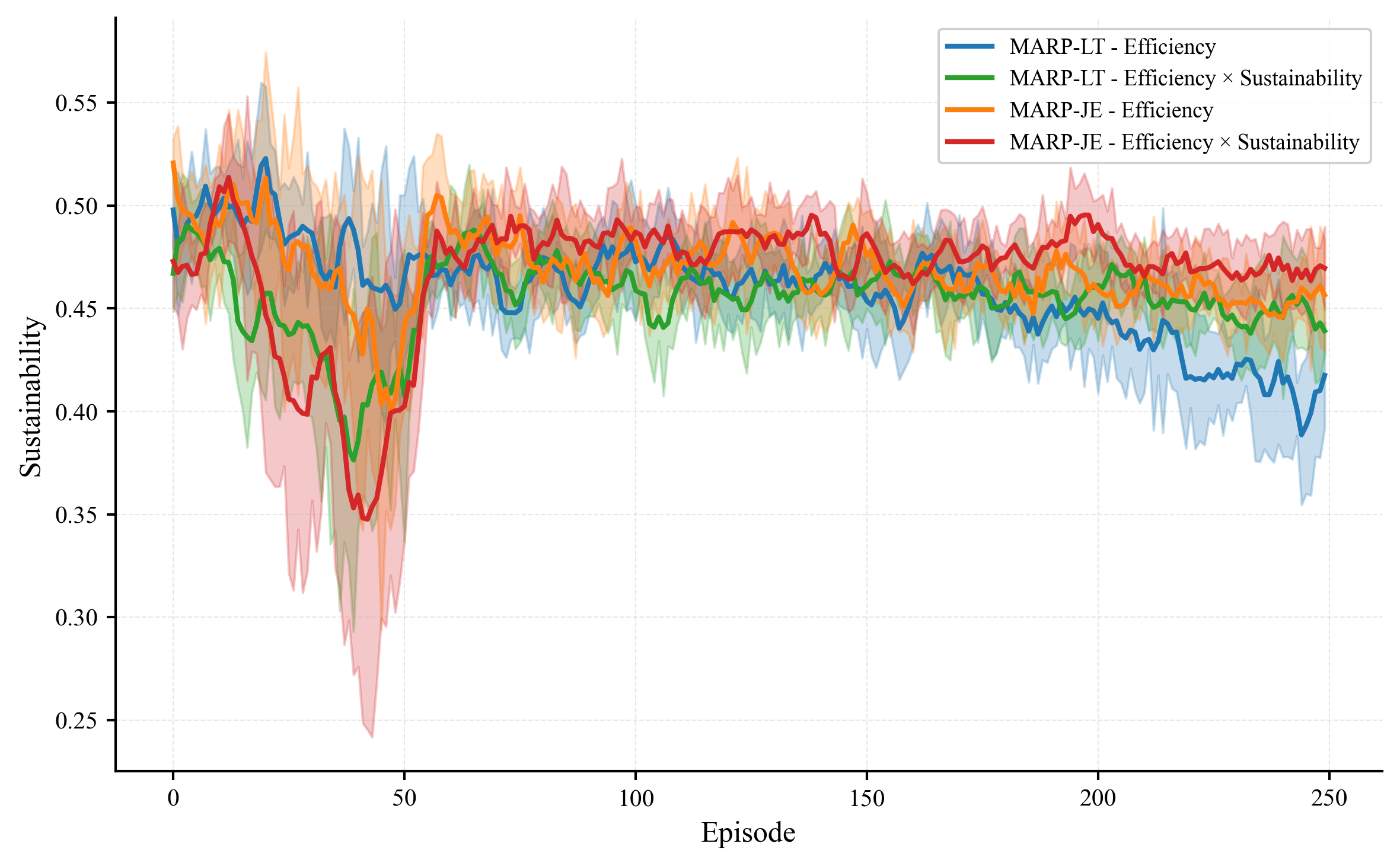}
        \caption{Sustainability}
        \label{fig:phi-b}
    \end{subfigure}

    \vspace{0.8em}

    \begin{subfigure}[t]{0.45\textwidth}
        \centering
        \includegraphics[width=\linewidth]{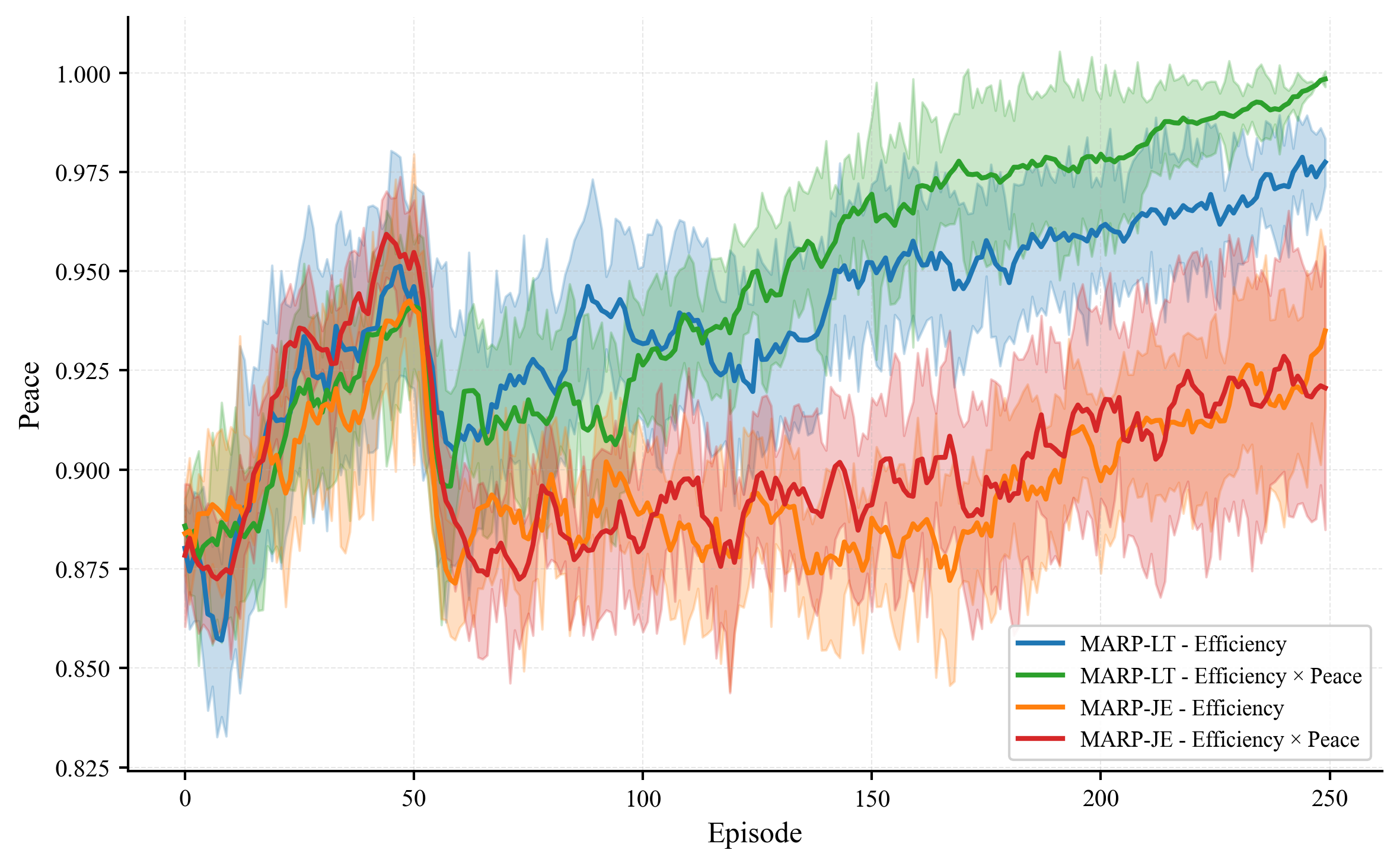}
        \caption{Peace}
        \label{fig:phi-c}
    \end{subfigure}

    \caption{For each social metric, we compare the two methods under two objectives: optimizing Efficiency alone and optimizing Efficiency combined with the respective metric. Results are averaged over five runs; error bars indicate standard error.}
    \label{fig:nv-ia-E-vs-dual}
\end{figure*}

This appendix provides additional empirical results that support the main text. Figure~\ref{fig:dqn-baseline} reproduces the \citet{perolat2017multi} setup using DQN with environment-defined rewards for comparison against our reward-prediction approach. Figure~\ref{fig:lt-reward} reports reward predictions under the Local-Trajectory variant across the three observation conditions discussed in Section~5.1. Figures~\ref{fig:social-metrics-all} and~\ref{fig:nv-ia-E-vs-dual} give the full breakdown of social metrics across single- and dual-objective training, complementing the summary in Section~5.2.

\FloatBarrier

\FloatBarrier

\end{document}